\documentclass[aps,pre,twocolumn,showpacs,superscriptaddress,groupedaddress]{revtex4}  
\usepackage{amsfonts}
\usepackage{amsmath}
\usepackage{amssymb}
\usepackage{version}
\usepackage{graphicx}%
\includeversion{New_connection}
\excludeversion{Old_connection}

\begin{document}
\title{Fluctuation-dissipation relation for systems with spatially varying friction}
\author{Oded Farago}
\affiliation{Department of Biomedical Engineering, Ben Gurion University of the Negev, Be'er Sheva, 84105 Israel}
\affiliation{Ilse Katz Institute for Nanoscale Science and Technology, Ben Gurion University of the Negev, Be'er Sheva, 84105 Israel}
\author{Niels Gr{\o}nbech-Jensen}
\affiliation{Applied Mathematics and Scientific Computing Department, Lawrence Berkeley National Laboratory, Berkeley, CA 94720}
\affiliation{Department of Mechanical and Aerospace Engineering, University of California, Davis, CA 95616}
\affiliation{Department of Chemical Engineering and Materials Science, University of California, Davis, CA 95616}

\begin{abstract}
When a particle diffuses in a medium with spatially dependent friction
coefficient $\alpha(r)$ at constant temperature $T$, it drifts toward
the low friction end of the system even in the absence of any real
physical force $f$. This phenomenon, which has been previously studied
in the context of non-inertial Brownian dynamics, is termed ``spurious
drift'', although the drift is real and stems from an inertial
effect taking place at the short temporal scales. Here, we study the
diffusion of particles in inhomogeneous media within the framework of the
inertial Langevin equation. We demonstrate that the quantity which
characterizes the dynamics with non-uniform $\alpha(r)$ is {\em
not}\/ the displacement of the particle $\Delta r=r-r^0$ (where $r^0$
is the initial position), but rather $\Delta A(r)=A(r)-A(r^0)$, where
$A(r)$ is the primitive function of $\alpha(r)$. We derive expressions
relating the mean and variance of $\Delta A$ to $f$, $T$, and the
duration of the dynamics $\Delta t$. For a constant friction
coefficient $\alpha(r)=\alpha$, these expressions reduce to the well
known forms of the force-drift and fluctuation-dissipation
relations. We introduce a very accurate method for Langevin dynamics
simulations in systems with spatially varying $\alpha(r)$, and use the
method to validate the newly derived expressions.
\end{abstract}
\maketitle

\section{Introduction}
\label{sec:intro}

In his ground-breaking 1905 paper on Brownian motion \cite{einstein},
Einstein noticed that the same random thermal forces from the
suspending medium that cause the diffusive motion of the particle,
also produce the friction experienced by the particle when pulled
through the same fluid medium.  From this observation, Einstein was
able to use statistical mechanics to derive the fluctuation-dissipation
relation
\begin{eqnarray}
D=\frac{k_BT}{\alpha},
\label{eq:fdr}
\end{eqnarray}
between the diffusion constant $D$, friction coefficient $\alpha$,
temperature $T$, and the Boltzmann constant $k_B$. Three years later,
Langevin introduced a very different approach to describe Brownian
motion \cite{langevin}. In contrast to Einstein, who considered the
Focker-Plank equation governing the particle's probability
distribution, Langevin focused on the particle's equation of motion
\begin{eqnarray}
m\dot{v}&=&f(r(t))-\alpha v+\beta(t),
\label{eq:lang}
\end{eqnarray}
where $m$ is the mass of the particle and $v(t)=\dot{r}$ is its
velocity. The Langevin equation describes Newtonian dynamics under the
influences of three forces: (i) a deterministic force $f$, (ii) a
friction force $-\alpha v$ proportional to the velocity with friction
coefficient $\alpha\ge0$, and (iii) a stochastic force $\beta(t)$
representing fluctuations arising from interactions with the embedding
medium that produces the friction. The stochastic force can be
conveniently modeled by a delta-correlated (``white'') Gaussian noise
with statistical properties \cite{parisi}:
\begin{eqnarray} 
\langle \beta(t)\rangle&=&0  \label{eq:noise1}\\
\langle\beta(t)\beta(t^\prime) \rangle&=&2\alpha k_BT\delta(t-t^\prime),
\label{eq:noise2}
\end{eqnarray}
where $\langle\cdots\rangle$ means a statistical average. This
statistical definition of $\beta(t)$ ensures that the motion satisfies
Einstein's fluctuation-dissipation relation, as can be easily realized
by considering the overdamped limit of Eq.~(\ref{eq:lang}) for a flat
potential ($f=0$). In this limit, the Langevin equation simplifies to
\begin{eqnarray}
\alpha v=\beta(t),
\label{eq:browneq}
\end{eqnarray}
and by integrating the equation over time and using
Eqs.~(\ref{eq:noise1}) and (\ref{eq:noise2}), one finds that the
displacement of the particle satisfies $\langle \Delta r\rangle=0$,
and
\begin{eqnarray}
\langle (\Delta r)^2\rangle=2Dt,
\label{eq:diffusion}
\end{eqnarray}
with $D=k_BT/\alpha$, as in Eq.~(\ref{eq:fdr}).

Langevin's work began a new field in mathematics that deals with
stochastic differential equations, namely equations in which one (or
more) of the terms is a stochastic process. Stochastic differential
equations require their own new calculus. The two most common versions
of stochastic calculi were proposed and developed by It\^{o} \cite{ito}
and Stratonovich \cite{stratonovich}. The difference between them
arises in equations with multiplicative noise, e.g., in
Eq.~(\ref{eq:browneq}) with a coordinate-dependent friction coefficient
$\alpha(r)$. In order to calculate the particle's trajectory, one needs
to integrate Eq.~(\ref{eq:browneq}) over a time interval $\Delta t$
\cite{doob}. For a uniform $\alpha$, the integral over the stochastic
noise is a well-defined Wiener process \cite{coffey,kampen}
\begin{eqnarray}
\int_0^{\Delta t} \beta(t) \, dt =\sqrt{2\alpha k_BT
\Delta t}\, \sigma,
\label{eq:weiner}
\end{eqnarray}
where $\sigma$ is a standard Gaussian random number satisfying
\begin{eqnarray}
\langle\sigma\rangle=0 \ \ ;\ \ \langle\sigma^2\rangle=1.
\label{eq:standard}
\end{eqnarray}
For a non-uniform $\alpha(r)$, the integral is ill-defined, since one
needs to specify both the trajectory and at which points along the
trajectory the friction coefficient in Eq.~(\ref{eq:weiner}) is
evaluated. In the It\^{o} convention \cite{ito}, the friction
coefficient is taken at the beginning of the time interval, while the
Stratonovich convention considers the algebraic mean of the initial
and final frictions \cite{stratonovich} (which, for a small time
interval, can be considered close to the friction at the
mid-point). Another commonly used convention is due to H\"{a}nggi
\cite{hanggi}. The latter uses the value of the friction coefficient
at the end of the time interval. In ordinary differential equations,
all the above conventions result in similar trajectories when the time
step becomes infinitesimal.  However, the Wiener process is
non-differentiable and, therefore, for the stochastic equation
(\ref{eq:browneq}), the different calculi lead to {\it different}\/
results of $r(t)$ for arbitrarily small integration time steps (see
Ref.~\cite{risken}, section 3.3.3).  The resulting ambiguity about the
appropriate way to interpret Eq.~(\ref{eq:weiner}) is known as the
It\^{o}-Stratonovich dilemma \cite{coffey,mannella}. Remarkably, it is
in fact the H\"{a}nggi convention that yields the correct equilibrium
distribution of the particle at constant $T$ \cite{lau}, which is the
reason why this interpretation is also known as the ``isothermal''
convention.

Diffusion in a medium with spatially dependent friction coefficient
raises yet another serious problem concerning the validity of the
fluctuation-dissipation theorem. Can one simply generalize
Eq.~(\ref{eq:fdr}) and write that $D(r)=k_BT/\alpha(r)$, and what is
the physical meaning of a coordinate dependent diffusion coefficient?
\cite{bhatt} The problem lies in the fundamental difference between
friction and diffusion. The former is a quantity that can be defined
locally by considering the motion of a particle in a flat potential at
zero temperature. Setting $f=0$ and $\beta=0$ in Eq.~(\ref{eq:lang}),
and integrating the equation over the time interval $\Delta t$, leads
to
\begin{eqnarray}
m\Delta v&=&-\int_0^{\Delta t}\alpha(r(t))v\, dt\label{eq:begin}
\\
&=&-\int_{r^0}^{r^0+\Delta
r}\alpha(r)\, dr
=-\alpha(r^0)\Delta r+{\cal
O}(\Delta r)^2,\nonumber
\end{eqnarray}
where $\Delta r=r(\Delta t)-r^0$ and $\Delta v=v(\Delta t)-v^0$ denote
the displacement and the change in velocity, respectively. Thus,
$\alpha(r^0)$ can be defined as the limiting value of $-m\Delta
v/\Delta r$, the ratio between the change in momentum and
displacement. In contrast to the friction coefficient, the diffusion
constant is not a local quantity, but is rather defined by the long
time asymptote of Eq.~(\ref{eq:diffusion}). As the particle diffuses
away from the point of origin, it explores new parts of the system and
experiences a varying friction coefficient. For a general friction
function $\alpha(r)$, it is not a-priory clear why the mean squared
displacement should even grow linearly with $t$, as implied by
Eq.~(\ref{eq:diffusion}). Recently, for instance, it has been argued
that certain functional forms of $\alpha(r)$ yield anomalous diffusion
where $\langle (\Delta r)^2\rangle\sim t^z$ with $z\neq 1$
\cite{metzler}. Moreover, the problem cannot be dealt with by
considering short time dynamics where the particle remains close to
the initial coordinate. Eq.~(\ref{eq:diffusion}) is relevant only on
time scales much larger than the relaxation time $\tau\sim m/\alpha$,
whereas on shorter time scales the motion of the particle is ballistic
and does not obey Eq.~(\ref{eq:diffusion}) at all (not even for a
uniform $\alpha$) \cite{coffey}.  These considerations suggest that
the concept of spatially dependent diffusion constant is somewhat
ambiguous, and that an alternative formulation for the
fluctuation-dissipation relation must be sought for.

In this paper we generalize the fluctuation-dissipation relation to
systems with non-uniform friction coefficients. The discussion extends
our previous study on the It\^{o}-Stratonovich dilemma, in which we
focused on the ``spurious drift'' (see section \ref{sec:spurforce}
below for an explanation of this term) of a particle in the presence of
a friction gradient \cite{prev}. Our treatment is based on the full
intertial Langevin equation (\ref{eq:lang}), and we highlight the fact
that the friction (dissipation) and noise (flucutuation) terms in this
equation are governed by slightly different friction coefficients.  We
reintroduce our new ``inertial'' convention, which has been developed
based on the analysis of Eq.~(\ref{eq:lang}), and which employs
different friction coefficients for the fluctuation and dissipation
contributions.  In Ref.~\cite{prev}, we found both the inertial and
isothermal conventions to produce the most accurate distribution
functions when implemented in Langevin dynamics simulations. Here, we
demonstrate that the former outperform the latter in cases when the
friction coefficient changes very rapidly. We use the simulations to
verify the validity of the newly derived fluctuation-dissipation
relationship, as well as of other theorertical predictions.

The paper is organized as follows: In section \ref{sec:theory} we
derive expressions for the drift (and the associated ``spurious
force'') experienced by a particle when traveling in a medium with
spatially dependent friction coefficient. We also present a generalized
form for the fluctuation-dissipation relation. The derived expressions
are tested and validated computationally in section
\ref{sec:simulations}, where we present our method for Langevin
dynamics simulations. The results are summarized and discussed in
section \ref{sec:summary}.

\section{Theory}
\label{sec:theory}

\subsection{The spurious force}
\label{sec:spurforce}

When a particle diffuses in a flat potential in a medium with constant
$\alpha$, the mean displacement of the particle (averaged over an
ensemble of stochastic trajectories or, equivalently, an ensemble of
particles) vanishes: $\langle \Delta r\rangle=0$. In the presence of a
friction gradient, the mean displacement does not vanish: $\langle
\Delta r\rangle\neq 0$ - a phenomenon that has been termed ``spurious
drift''. The drift, which is in the opposite direction to the friction
gradient, is, of course, not spurious, but rather represents the effect
of inertia.  It originates from the fact that when the particle
travels toward a less viscous regime (i.e., against the friction
gradient), it suffers less dissipation and therefore travels longer
distances. This inertial effect is countered by a ``trapping effect''
that takes place on time scales larger than the ballistic relaxation
time $\tau\sim m/\alpha$, and which has precisely the same origin,
namely the fact that the ballistic distance decreases with
$\alpha$. At the large time scales, the larger friction slows down the
diffusion of the particle and, thus, traps it in the more viscous
regime. In the case of a flat potential, the equilibrium distribution
of the particle [which is independent of $\alpha(r)$] is uniform,
which means that, on average, the particle spends the same amount of
time in each part of the system. This implies that the drift, which
favors the low viscosity regime, precisely balances the slower
diffusion on the high viscosity end.

Since the drift is an inertial effect, it must be dealt with within
the framework of the full inertial Langevin equation (\ref{eq:lang}),
and not by using its strictly overdamped, non-inertial form
Eq.~(\ref{eq:browneq}). Assuming a flat potential ($f=0$), and
integrating Eq.~(\ref{eq:lang}) over the time interval from $t=0$ to
$t=\Delta t$, we arrive at the ``integrated Langevin equation''
\begin{eqnarray}
m\Delta v=-\int_{r^0}^{r^0+\Delta
r}\alpha(r)\, dr+\int_0^{\Delta t}\beta(t)\, dt.
\label{eq:intlang}
\end{eqnarray}
The terms on the r.h.s.\ of Eq.~(\ref{eq:intlang}), which give the
friction and noise contributions to the change in the momentum, are
governed by two distinct friction coefficients representing different
averages of the friction function during the time interval. The
friction term features the spacially averaged friction coefficient,
$\bar{\alpha}_r$
\begin{eqnarray}
\bar{\alpha}_r&=&\frac{\int_{r^0}^{r^0+\Delta
r}\alpha(r)\, dr}{\Delta r}=\frac{A(r^0+\Delta
r)-A(r^0)}{\Delta r}\equiv\frac{\Delta A}{\Delta r},\nonumber \\
&&
\label{eq:rfriction}
\end{eqnarray}
where $A(r)$ is the primitive function of $\alpha(r)$.  The
spatially averaged friction coefficient, $\bar{\alpha}_r$, has the
following properties: (i) It depends on the initial and final
coordinates, but not on the trajectory $r(t)$ between them. (ii) It is
a well defined quantity that exists even if $\alpha(r)$ is
discontinuous. (iii) For smooth friction functions and sufficiently
small $\Delta r$, $\bar{\alpha}_r$ is well approximated by the Stratonovich
friction coefficient
\begin{eqnarray}
\bar{\alpha}_r\simeq \frac{\alpha(r^0)+\alpha(r^0+\Delta r)}{2}\simeq
\alpha(r^0)+\frac{\alpha^{\prime}(r^0)}{2}\Delta r.
\label{eq:expansion}
\end{eqnarray}
The noise term in Eq.~(\ref{eq:intlang}) is governed by a different
friction coefficient. This term represents a sum of random Gaussian
variables with vanishing correlation time. Therefore, it can be {\em
formally}\/ written as an integral of a Wiener process [compare with
Eq.~(\ref{eq:weiner})]
\begin{eqnarray}
\int_0^{\Delta t} \beta(t) \, dt =\sqrt{2\bar{\alpha}_t k_BT
\Delta t}\, \sigma,
\label{eq:illweiner}
\end{eqnarray}
where $\sigma$ is a standard Gaussian random variable [see
Eq.~(\ref{eq:standard})], and
\begin{eqnarray}
\bar{\alpha}_t=\frac{\int_{0}^{\Delta
t}\alpha(r(t))\, dt}{\Delta t}
\label{eq:tfriction}
\end{eqnarray}
is the time-averaged friction coefficient. In contrast to
$\bar{\alpha}_r$, which depends only on the end points of the
interval, the calculation of $\bar{\alpha}_t$ requires full knowledge
of the trajectory, $r(t)$, during the time step. Without this
information, $\bar{\alpha}_t$ cannot be uniquely determined since
there exists not just one path, but a distribution of possible
trajectories, leading from $r^0$ to $r^0+\Delta r$. This is the origin
of the It\^{o}-Stratonovich dilemma. While Eq.~(\ref{eq:tfriction}) is
formally correct, it bears no physical meaning for non-zero time steps
as it is based on the assumption that the noise is temporally
uncorrelated (white), which is only true for vanishing $\Delta t$. For
time steps $\Delta t>0$, the friction gradient colors the noise, since
the noise value at one time instance changes the trajectory of the
particle and, thereby, influences the noise statistics at a following
instance in time.

To address the above problem and make Eq.~(\ref{eq:intlang})
physically unambiguous, we need to consider the ensemble average over
all possible trajectories starting at $r^0$, rather than a single path
of the particle. On the l.h.s.\ of Eq.~(\ref{eq:intlang}) we have the
change in the momentum of the particle which, in the absence of
deterministic forces ($f=0$), must have a vanishing ensemble
average. On the r.h.s.\ we have the friction and noise forces. In
accordance with Einstein's idea, these terms arise from the random
forces caused by the collisions with the molecules of the thermal
bath. The collisions occur at such a fast rate that it can be assumed
that the particle barely moves before experiencing enough collisions
to make the central limit theorem applicable (see discussion in
section 4.5 of Ref.~\cite{gillespie}). Thus, the total change in the
momentum of the particle during an arbitrarily small time step is
normally distributed.  The friction force represents the mean rate of
change in the momentum, while the noise accounts for the statistical
fluctuations around the mean and, therefore, has a vanishing ensemble
average at each instance during the time interval $\Delta t$. This
implies that the ensemble average of the change in the momentum due to
the noise must vanish
\begin{eqnarray}
\left\langle\int_0^{\Delta t} \beta(t) \, dt \right\rangle=0,
\label{eq:zeronoise}
\end{eqnarray}
and this feature must be incorporated in the integral form of the
Langevin equation (\ref{eq:intlang}) to make it consistent with the
fluctuation-dissipation relationship. That leaves us with only the
friction term in Eq.~(\ref{eq:intlang}) whose ensemble average must
therefore also vanish, and by using Eq.~(\ref{eq:rfriction}) we
conclude that the drift satisfies
\begin{eqnarray}
\langle \bar{\alpha}_r\Delta r\rangle=\langle\Delta A\rangle=0.
\label{eq:drift}
\end{eqnarray}
Notice that Eq.~(\ref{eq:drift}) holds for any time interval $\Delta
t$ (i.e., both in the ballistic and diffusive regimes), for as long as
$f=0$. For a constant $\alpha$ it reduces to the no-drift condition:
$\langle\Delta r\rangle=0$.

As noted above, the drift does not arise from the action of any real
force but rather from the friction gradient.  Using
Eq.~(\ref{eq:expansion}) in (\ref{eq:drift}) we arrive at
\begin{eqnarray}
\langle \Delta r\rangle \simeq -\frac{\alpha^{\prime}}{2\alpha}
\langle(\Delta r)^2\rangle.
\label{eq:fricgrad}
\end{eqnarray}
The associated spurious force is defined as the force generating a
similar drift in a uniform medium. At short time scales, $\Delta t \ll
\tau\sim m/\alpha$, the motion of the particle is ballistic ($\Delta
r\simeq v\Delta t$) and, thus, $\langle (\Delta r)^2\rangle\simeq
\langle (v\Delta t)^2\rangle =(k_BT/m)\Delta t^2$, where the second
equality is obtained by virtue of the equilibrium Maxwell-Boltzmann
velocity distribution. We thus conclude that in the ballistic regime,
the drift is given by
\begin{eqnarray}
\langle \Delta r\rangle\simeq -\frac{1}{2}\frac{\alpha^\prime}{\alpha}
\frac{k_BT}{m}\Delta t^2,
\end{eqnarray}
which resembles the inertial Newtonian dynamics of a particle under
the action of a (spurious) force
\begin{eqnarray}
f_s=-k_BT \left(\frac{\alpha^{\prime}}{\alpha}\right).
\label{eq:spuriousf1}
\end{eqnarray}

On time scales much larger than the ballistic correlation time,
$\Delta t\gg \tau$, the motion must be compared to the diffusive
dynamics of a particle in a uniform medium. For such a particle
$\langle (\Delta r)^2\rangle=2D\Delta t=2(k_BT/\alpha)\Delta t$, and
by inserting this relationship into Eq.~(\ref{eq:fricgrad}), we arrive
at the following expression for the ``spurious velocity'',
$v_s\equiv\langle \Delta r\rangle/\Delta t$
\begin{eqnarray}
\alpha v_s=-k_BT \left(\frac{\alpha^{\prime}}{\alpha}\right).
\end{eqnarray}
This result can be compared to the velocity of a particle dragged by a
(spurious) force of magnitude
\begin{eqnarray}
f_s=-k_BT \left(\frac{\alpha^{\prime}}{\alpha}\right),
\label{eq:spuriousf2}
\end{eqnarray}
in a liquid with friction coefficient $\alpha$. Remarkably,
Eqs.~(\ref{eq:spuriousf1}) and (\ref{eq:spuriousf2}) provide identical
expressions for the spurious force at both the short- and long-time
limits, representing both ballistic and diffusive behavior. Notice
that the spurious force calculation derived here is based on a first
order expansion of $\alpha(r)$ in $\Delta r$
[Eq.~(\ref{eq:expansion})], which is only valid for smoothly varying
friction and requires that the variation in $\alpha(r)$ during the
time interval is relatively small ($\alpha^{\prime}\Delta r \ll
\alpha$, i.e., a weak spurious force). This is a reasonable
approximation in the short-time ballistic regime, but may be
questionable in the diffusive time-scale. In what follows (see
especially section \ref{sec:dfr} and the results in
Fig.~\ref{fig:step}), we will go beyond this approximation and allow
larger variations in $\alpha(r)$.

\subsection{Force measurements}

Part of the renewed interest in the It\^{o}-Stratonovich dilemma stems
from the relevance of the topic to experiments involving femto-Newton
force measurements \cite{lancon,volpe}. When the particle under
investigation is found close to the surface of the sample cell, its
diffusion coefficients parallel and perpendicular to the boundary
decrease due to the hydrodynamic interactions between the surface and
particle. In light of the debate that has erupted about the
interpretation of the results of such experiments
\cite{volpe,mannella2}, we here use our formalism to derive a new
expression relating the displacement and the deterministic force
acting on the particle. Since it is impossible to address the force
variations on time scales smaller than the measurement interval
$\Delta t$, we will assume that the force $f$ is constant during this
time frame. Without any further assumptions, we start with
Eq.~(\ref{eq:intlang}), but now in the presence of a constant
deterministic force $f$, and we take the ensemble average of the
different terms. Using Eqs.~(\ref{eq:rfriction}) and
(\ref{eq:zeronoise}) we arrive at
\begin{eqnarray}
m\langle \Delta v\rangle=f\Delta t-\langle  \bar{\alpha}_r\Delta r\rangle
=f\Delta t-\langle  \Delta A\rangle.
\label{eq:intlang2}
\end{eqnarray}
In section \ref{sec:spurforce} we argued that for $f=0$, there can be
no change in the (ensemble) average momentum of the particle, which
means that the l.h.s.\ of Eq.~(\ref{eq:intlang2}) must vanish for any
$\Delta t$. This is obviously not true when $f\neq 0$ since the force
results in a change in the momentum. However, when $\Delta t\gg \tau$,
the velocity of the particle becomes uncorrelated with the initial
velocity at $t=0$. Starting at $r^0$ with an ensemble of particles
with an equilibrium velocity (Maxwell-Boltzmann) distribution, the
velocity distribution at $\Delta t\gg\tau$ is expected to attain the
same equilibrium form. Therefore, for $\Delta t\gg \tau$, the result
$\langle \Delta v\rangle=0$ still holds, and when used in
Eq.~(\ref{eq:intlang2}) it leads to
\begin{eqnarray}
f=\frac{\langle \bar{\alpha}_r\Delta r\rangle}{\Delta t}=\frac{\langle
\Delta A\rangle}{\Delta t}.
\label{eq:forcemes}
\end{eqnarray}
When the friction coefficient is constant, this expression becomes
$f=\alpha\langle v\rangle$, where $\langle v\rangle\equiv\langle\Delta
r\rangle/\Delta t$ is the drift velocity.

When the variation in $\alpha(r)$ during the time interval is small (yet
non-negligible!), the truncated expansion on the r.h.s.\ of
Eq.~(\ref{eq:expansion}) can be used in Eq.~(\ref{eq:forcemes}),
yielding
\begin{eqnarray}
f\simeq \frac{\alpha\langle \Delta r\rangle}{\Delta
t}+\frac{\alpha^{\prime}\langle (\Delta r)^2\rangle}{2\Delta t}.
\label{eq:volpe1}
\end{eqnarray}
If we {\em define}\/ a spatially dependent diffusion coefficient as
$D(r)=k_BT/\alpha(r)$, and also assume that $\langle (\Delta
r)^2\rangle=2D\Delta t$, then Eq.~(\ref{eq:volpe1}) can be rewritten as 
\begin{eqnarray}
f\simeq \frac{\alpha\langle \Delta r\rangle}{\Delta
t}-\alpha D^{\prime},
\label{eq:volpe2}
\end{eqnarray}
where $D^{\prime}\equiv dD/dr=-k_BT(\alpha^{\prime}/\alpha^2)$.  The
last equation has been used in Ref.~\cite{volpe} for the force
measurements. Unlike Eq.~(\ref{eq:forcemes}) which is asymptotically
correct (for $\Delta t\gg\tau$), Eq.~(\ref{eq:volpe2}) is an
approximation involving a first order expansion of $\alpha(r)$ in
$\Delta r$ [Eq.~(\ref{eq:expansion})]. This approximation is valid
only when $\alpha(r)$ is a well-behaved smooth function and the
variations in $\alpha(r)$ during the time interval $\Delta t$ are
small. Additionally, Eq.~(\ref{eq:volpe2}) also features a spatially
dependent diffusion coefficient $D(r)=k_BT/\alpha(r)$, and involves
the assumption that $\langle (\Delta r)^2\rangle=2D\Delta t$. The
latter result is an approximate form of the fluctuation-dissipation
relation, which is correct only if the variations in $\alpha(r)$ are
neglected to zeroth order [i.e., assuming $\alpha=\alpha(r^0)$]. In
the following section \ref{sec:dfr} we present a generalized form for
the fluctuation-dissipation relationship that takes the variations in
$\alpha(r)$ into account. In section \ref{sec:fick} we discuss the
physical meaning of a spatially dependent diffusion coefficient in
general, and the definition $D(r)=k_BT/\alpha(r)$ in particular.

\subsection{The fluctuation-dissipation relationship}
\label{sec:dfr}

To derive the correct form of the fluctuation-dissipation
relationship, we set $f=0$, and start with Eq.~(\ref{eq:intlang}),
which we now write in a slightly different form
\begin{eqnarray}
m\Delta v+\bar{\alpha}_r\Delta r=\int_0^{\Delta t}\beta(t^{\prime})\, dt^{\prime}.
\label{eq:intlang3}
\end{eqnarray}
By squaring the equation and taking the ensemble average, we arrive at 
\begin{eqnarray}
\left\langle \left(m\Delta v\right)^2+2m\Delta v\bar{\alpha}_r\Delta r
+\left(\bar{\alpha}_r\Delta r\right)^2\right\rangle\nonumber \\
=\int_0^{\Delta t}\!\!dt^{\prime\prime}\int_0^{\Delta t}\!\!dt^{\prime}\,
\left \langle\beta(t^{\prime\prime})\beta(t^{\prime})\right\rangle.
\label{eq:fdr1}
\end{eqnarray}
At large times, $\Delta t\gg\tau$, the expression on the l.h.s.\ of
this equation is dominated by the third term, which roughly grows
linearly with $\Delta t$ while the other two terms remain finite. The
term on the r.h.s.\ can be evaluated by using Eq.~(\ref{eq:noise2})
with $\alpha=\alpha(t)$. This leads us to the asymptotic equation
\begin{eqnarray}
\left\langle \left(\bar{\alpha}_r\Delta
r\right)^2\right\rangle=\left\langle \left(\Delta
A\right)^2\right\rangle =\int_0^{\Delta t}
2\langle\alpha(t)\rangle k_BT\, dt,
\label{eq:fdrfinal}
\end{eqnarray}
which is the generalized form of the fluctuation-dissipation
relationship for system with spatially dependent friction.  For a constant
friction coefficient $\alpha$, the relationship reduces to the well
known form $\langle(\Delta r)^2\rangle=2(k_BT/\alpha)\Delta t$. Notice
that $\langle\alpha(t)\rangle=\langle\alpha(r(t))\rangle$ can also be
expressed as
\begin{eqnarray}
\langle\alpha(r(t))\rangle=\int_0^{\Delta t}\rho(r,t)\alpha(r)\, dr,
\label{eq:avalphat}
\end{eqnarray}
where $\rho(r,t)$ is the normalized distribution function of the
particle at time $t$. This implies that $\langle\alpha(r(t))\rangle$
depends on the initial distribution $\rho(r,0)$. If the particle is
initially localized at $r=r^0$, then
$\rho(r,0)=\delta(r-r^0)$.

Another interesting case is that of an infinite system with average
nonzero density $\rho_0=1/L$, modeled by periodic boundary conditions
to a system with finite length $L$. If the initial distribution
$\rho(r,0)$ coincides with the equilibrium distribution, which (for
$f=0$) is uniform $\rho(r,t)=\rho_{\rm eq}(r)=1/L$, then
Eq.~(\ref{eq:fdrfinal}) simplifies to
\begin{equation}
\left\langle \left(\Delta
A\right)^2\right\rangle =2\langle\alpha\rangle k_BT\Delta t,
\label{eq:fdrspecial}
\end{equation}
with $\langle\alpha\rangle=L^{-1}\int \alpha(r) dr$.  However, the far
l.h.s.\ of Eq.~(\ref{eq:fdrfinal}) {\em cannot}\/ be replaced with
$\langle\alpha\rangle^2\langle(\Delta r)^2\rangle$ [or
$\langle\alpha^2\rangle\langle(\Delta r)^2\rangle$] in this case,
since the latter form does not account correctly for the drift of the
particle. This highlights the fact that $\Delta A$, and not $\Delta
r$, is the quantity that characterizes the displacement of the
particle when traveling in an inhomogeneous medium. This conclusion is
also reflected in Eqs.~(\ref{eq:drift}) and (\ref{eq:forcemes}).

\subsection{Fick's second law}
\label{sec:fick}

We conclude the analytical part of the paper by returning to our
earlier comment [see text after Eq.~(\ref{eq:begin})] that the concept
of a spatially dependent friction coefficient $D(r)$ is non-trivial
since diffusion is inherently a non-local process. This has motivated
us, throughout section \ref{sec:theory}, to derive expressions
involving only the local friction coefficient $\alpha(r)$. The only
context in which $D(r)$ can be rationalized is the Focker-Planck
equation for the probability density of the particle $\rho(r,t)$,
which can be derived as follows: For the simplicity of the
presentation (but without limiting the generality of the derived
equation), let us assume that the particle initially is located at
$r^0$ [i.e., $\rho(r,0)=\delta(r-r^0)$]. We start by rewriting
Eq.~(\ref{eq:fdrfinal}) [together with Eq.~(\ref{eq:avalphat})] in the
following explicit form
\begin{equation}
\int_{-\infty}^{\infty}\!\!\!dr^{\prime}
B^2(r^{\prime})\rho(r^{\prime},t)
=2k_BT\!\int_0^t
\!\!\!dt^{\prime}\!\int_{-\infty}^{\infty}\!\!\!dr^{\prime}\alpha(r^{\prime})\rho(r^{\prime},t^{\prime}),
\end{equation}
where $B(r)\equiv\Delta A(r)=A(r)-A(r^0)$. Taking the partial
derivative with respect to $t$ gives
\begin{eqnarray}
&&\int_{-\infty}^{\infty}\!\!\!dr^{\prime} 
B^2(r^{\prime})\frac{\partial \rho(r^{\prime},t)}
{\partial t}
=2k_BT
\!\int_{-\infty}^{\infty}\!\!\!dr^{\prime}\alpha(r^{\prime})\rho(r^{\prime},t)\nonumber \\
&&=-2k_BT
\!\int_{-\infty}^{\infty}\!\!\!dr^{\prime}B(r^{\prime})\frac{\partial \rho(r^{\prime},t)}{\partial r^{\prime}},
\label{eq:fick1}
\end{eqnarray}
where the second equality is obtained via integration by parts,
keeping in mind that $B^{\prime}(r)=\alpha(r)$ and using the fact that
$\rho(r,t)$ vanishes for $r\rightarrow\pm\infty$. By multiplying and
dividing the integrand on the r.h.s.\ of Eq.~(\ref{eq:fick1}) by
$\alpha(r^{\prime})$, and using the identify
$2B(r)\alpha(r)=2B(r)B^{\prime}(r)=[B^2(r)]^{\prime}$, we arrive at
\begin{eqnarray}
&&\int_{-\infty}^{\infty}\!\!\!dr^{\prime}
B^2(r^{\prime})\frac{\partial
\rho(r^{\prime},t)}{\partial t} =\nonumber \\
&&-k_BT
\!\int_{-\infty}^{\infty}\!\!\!dr^{\prime}\left[B^2(r^{\prime})\right]^{\prime}\frac{1}{\alpha(r^{\prime})}\frac{\partial
\rho(r^{\prime},t)}{\partial r^{\prime}}.
\label{eq:fick2}
\end{eqnarray}
Integrating by parts the r.h.s.\ of Eq.~(\ref{eq:fick2}) yields
\begin{eqnarray}
&&\int_{-\infty}^{\infty}\!\!\!dr^{\prime} B^2(r^{\prime})
\frac{\partial
\rho(r^{\prime},t)}{\partial t} =\nonumber \\
&&\!\!\!\int_{-\infty}^{\infty}\!\!\!dr^{\prime}B^2(r^{\prime})\frac{\partial}{\partial
r^{\prime}}\left[\frac{k_BT}{\alpha(r^{\prime})}\frac{\partial \rho(r^{\prime},t)}{\partial
r^{\prime}}\right].
\label{eq:fick3}
\end{eqnarray}
Since Eq.~(\ref{eq:fick3})  holds for any function $B(r)$ it must be that
\begin{equation}
\frac{\partial \rho(r,t)}{\partial
t}=\frac{\partial}{\partial
r}\left[\frac{k_BT}{\alpha(r)}\frac{\partial
\rho(r,t)}{\partial r}\right].
\label{eq:ficklaw}
\end{equation}
The last equation is Fick's second law, which is commonly written as
$\partial_t \rho=\partial_r[D(r)\partial_r\rho]$, with $D(r)$ being
the spatially dependent diffusion coefficient. Comparing this form to
Eq.~(\ref{eq:ficklaw}), we find that $D(r)=k_BT/\alpha(r)$, which is
the natural generalization of Eq.~(\ref{eq:fdr}).

\section{Langevin dynamics simulations}
\label{sec:simulations}

In the previous section we demonstrated that much of the
It\^{o}-Stratonovich dilemma can be resolved by: (i) considering the
inertial Langevin equation (\ref{eq:lang}) rather than its overdamped,
non-inertial limit (\ref{eq:browneq}), (ii) taking the ensemble
average over many stochastic trajectories, and (iii) enforcing
Eq.~(\ref{eq:zeronoise}) for the contribution of the noise to the
momentum of the particle. This has led to the derivation of
Eqs.~(\ref{eq:drift}), (\ref{eq:forcemes}), and (\ref{eq:fdrfinal}),
which we now test using computer simulations.  When performing
Langevin dynamics simulations, a set of algebraic equations (an
``integrator'') is used to generate stochastic trajectories of the
particle. Choosing the appropriate convention to be implemented in the
integrator invokes the It\^{o}-Stratonovich dilemma in a slightly
different form, as will be discussed in the following section.

For the Langevin dynamics simulations, we use the GJF integrator
\cite{gjf} which, starting with $r=r^n$ and $v=v^n$ at $t=t_n$, uses
the following equations for calculating the position, $r^{n+1}$, and
velocity, $v^{n+1}$, at time $t_{n+1}=t_n+dt$
\begin{eqnarray}
\!\!\!\!r^{n+1}\!&=&r\!^n+bdtv^n+\frac{bdt^2}{2m}f^n
+\frac{bdt}{2m}\sqrt{2\alpha k_BTdt}\,\sigma^{n+1}\nonumber
\\
\label{eq:pos-equation}\\
\!\!\!\!v^{n+1}\!&=&\!av^n\!+\!\frac{dt}{2m}\left(af^n+f^{n+1}\right)\!
+\!\frac{b}{m}\sqrt{2\alpha
k_BTdt}\,\sigma^{n+1}, \nonumber \\
\label{eq:vel-equation}
\end{eqnarray}
where $f^n=f(r^n)$, $\sigma^n$ is a random Gaussian number satisfying
Eq.~(\ref{eq:standard}), and the coefficients $a$ and $b$ are given by
\begin{eqnarray}
b&=&\left(1+\frac{\alpha  dt}{2m}\right)^{-1} \label{eq:bcoef}
\\
a&=&b\left(1-\frac{\alpha  dt}{2m}\right).
\label{eq:acoef}
\end{eqnarray}

For a constant friction coefficient $\alpha$, it was analytically
demonstrated that the GJF integrator provides exact thermodynamic
response for both flat and harmonic potentials for any time step $dt$
within the stability criterion of the method~\cite{gjf}. For spatially
dependent friction $\alpha(r)$, one needs to choose the value of
$\alpha$ to be used in
Eqs.~(\ref{eq:pos-equation})-(\ref{eq:acoef}). The conventions of
It\^{o}, Stratonovich, and (H\"{a}nggi) (the isothermal) correspond to
setting $\alpha=\alpha(r^n)$,
$\alpha=[\alpha(r^n)+\alpha(r^{n+1})]/2$, and
$\alpha=\alpha(r^{n+1})$, respectively. None of these interpretations
is physically accurate since, as our discussion in section
\ref{sec:spurforce} reveals, the important friction coefficients are
$\bar{\alpha}_r$ (\ref{eq:rfriction}) and $\bar{\alpha}_t$
(\ref{eq:tfriction}). The former governs the friction term in
Eq.~(\ref{eq:intlang}) and, therefore, is the one to be used in
expressions (\ref{eq:bcoef}) and (\ref{eq:acoef}) for the coefficients
$b$ and $a$ characterizing the dissipation decay rate of the
velocity. The latter should be used for the noise amplitude,
$(2k_B\bar{\alpha}_tdt)^{1/2}$. For smooth friction functions,
$\bar{\alpha}_r$ differs by ${\cal O}(dt)$ from the value used in the
It\^{o} and isothermal interpretations, and by ${\cal O}(dt^2)$ from
the Stratonovich value. This may indicate that the most accurate
interpretation is that of Stratonovich. Unfortunately, the
Stratonovich friction coefficient uses information about the position
of the particle at the end of the time step. Therefore, using this
value in Eq.~(\ref{eq:illweiner}), would result in violation of
Eq.~(\ref{eq:zeronoise}), which must be satisfied by the stochastic
noise term. The isothermal interpretation suffers from exactly the
same deficiency of the noise term, while It\^{o}'s convention, despite
satisfying Eq.~(\ref{eq:zeronoise}), assumes a value which clearly
deviates by ${\cal O}(dt)$ from $\bar{\alpha}_t$.

In our previous work we proposed a new ``inertial'' convention
\cite{prev}, where $\bar{\alpha}_r$ (\ref{eq:rfriction}) is used for
the coefficients $a$ and $b$, while
\begin{eqnarray} 
\bar{\alpha}_t&\simeq&\bar{\alpha}_r(r^n\to r^n+v^n dt)=
\frac{\int_{r^n}^{r^n+v^n dt}\alpha(r)\, dr}{v^n dt}\nonumber \\
&=&\alpha(r^n)+\alpha^{\prime}(r^n)\frac{v^ndt}{2}
+{\cal O}(dt^2),
\label{eq:alphat1}
\end{eqnarray}
is used for the noise amplitude. The fact that the friction
coefficient given by Eq.~(\ref{eq:alphat1}) is based on information
existing at $t=t_n$ only, makes $(2\bar{\alpha}_tk_BTdt)^{1/2}\sigma$
a true Gaussian variable and ensures that Eq.~(\ref{eq:zeronoise}) is
obeyed. Expression (\ref{eq:alphat1}) is essentially the best guess
that one can make for $\bar{\alpha}_t$ at $t=t_n$. It involves the
assumption that the particle travels with velocity $v^n$ during the
time step. This is a reasonable estimation of $\bar{\alpha}_t$ for
small time steps $dt\ll\tau$, during which the trajectory of the
particle is nearly ballistic.

\begin{figure}[t]
\begin{center}
\scalebox{0.365}{\centering \includegraphics{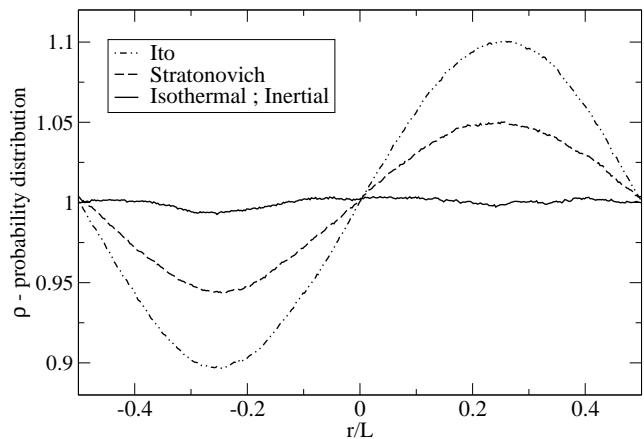}}
\end{center}
\vspace{-0.5cm}
\caption{The probability distribution computed using It\^{o}
(dashed-dotted), Stratonovich (dashed), isothermal (solid line), and
the new inertial (same solid line - curve is indistinguishable from
the isothermal curve at the resolution of the graph)
interpretations. The results correspond to a particle diffusing in a
flat potential with a friction function that has a sinusoidal
form. The results for the different conventions were computed with the
same normalized time step $dt=0.1$.}
\label{fig:smooth}
\end{figure}

Obviously, neither the newly proposed inertial convention nor the
above mentioned more familiar ones (It\^{o}, Stratonovich, isothermal)
are exact for discrete time steps. The fact that the integrator
numerically solves the inertial Langevin equation (\ref{eq:lang}) and
not its non-inertial form (\ref{eq:browneq}), guarantees that the
correct equilibrium distribution is obtained when $dt\rightarrow 0$
for {\it any} sensible interpretation. The difference between the
conventions, as implemented for inertial Langevin dynamics, is simply
the rate of convergence to the correct distribution for
$dt\rightarrow0$. This may seem as a lighter version of the
It\^{o}-Stratonovich dilemma, which is {\it only} a fundamental issue
if the inertial term is entirely omitted in the Langevin
equation. However, the rate of convergence has a considerable
practical importance in simulations where the time step $dt$ is not
infinitesimal. The difference between the conventions is demonstrated
in Fig.~\ref{fig:smooth}, showing the simulated spatial equilibrium
distribution of a particle of normalized mass $m=1$, in contact with a
constant temperature bath $T$, moving in a one-dimensional medium with
a flat potential and a sinusoidal normalized friction coefficient
given by $\alpha(r)=2.75+2.25\sin(2\pi r/L)$, where $L=40$ is the
spatial extension of the system in normalized units. All the results
depicted in Fig.~\ref{fig:smooth} were derived from simulations with
normalized time step $dt=0.1$. Since the potential energy is constant,
the equilibrium distribution must be uniform. Our results show that
both It\^{o} and Stratonovich interpretations exhibit noticeable
deviations from the correct uniform equilibrium distribution. The
deviations reflect the sinusoidal form of the friction function. In
contrast, the isothermal and inertial conventions produce
indistinguishable distributions that are fairly uniform and deviate by
less than $0.5\%$ from the correct value of 1.

The ability of the isothermal and inertial conventions to accurately
sample the equilibrium distribution function while using relatively
large time steps was discussed in details in Ref.~\cite{prev}. In
short, the reason lies in the fact that these conventions account
correctly for the drift of the particle, although this happens in very
different ways. In the inertial convention the drift originates from
the dissipation term in the integrated Langevin equation, while the
noise term in that equation has zero mean, in accordance with
Eq.~(\ref{eq:zeronoise}). In contrast, in the isothermal convention,
Eq.~(\ref{eq:zeronoise}) for the noise is not satisfied, and the drift
is generated by the friction term being larger than
necessary. Fortunately for the isothermal convention, these two errors
cancel each other. The deviations from a uniform probability
distribution of the It\^{o} and Stratonovich conventions originate
from ${\cal O}(dt^2)$ errors in the computed drift, which can be
corrected at the end of each time-step by adding a ``spurious drift
term'' to $r^{n+1}$. Based on this strategy, we presented yet another
convention, the ``corrected-Stratonovich convention'', in
Ref.~\cite{prev}. While for smooth friction functions, the latter
convention is computationally almost as good as the isothermal and
inertial conventions, it is not useful for systems where $\alpha(r)$
exhibit rapid spatial variations. We, therefore, focus on a comparison
between the isothermal and inertial conventions. In the simulations
reported in Fig.~\ref{fig:smooth}, we observed that, when starting
with the same initial position and velocity and using the same seed
for the Gaussian random number generator, the isothermal and inertial
conventions produced nearly identical trajectories, which explains why
the resulting probability distributions depicted in
Fig.~\ref{fig:smooth} are indistinguishable.

The ability of the isothermal and inertial conventions to accurately
sample the equilibrium distribution function while using relatively
large time steps was discussed in details in Ref.~\cite{prev}. In
short, the reason lies in the fact that these conventions account
correctly for the drift of the particle, although this happens in very
different ways. In the inertial convention the drift originates from
the dissipation term in the integrated Langevin equation, while the
noise term in that equation has zero mean, in accordance with
Eq.~(\ref{eq:zeronoise}). In contrast, in the isothermal convention,
Eq.~(\ref{eq:zeronoise}) for the noise is not satisfied, and the drift
is generated by the friction term being larger than necessary. Fortunately
for the isothermal convention, these two errors cancel each other. In
simulations we observed that, when starting with the same initial
position and velocity and using the same seed for the Gaussian random
number generator, the isothermal and inertial conventions produced
nearly identical trajectories, which explains why the resulting
probability distributions depicted in Fig.~\ref{fig:smooth} are
indistinguishable.

\begin{figure}[t]
\begin{center}
\scalebox{0.55}{\centering \includegraphics{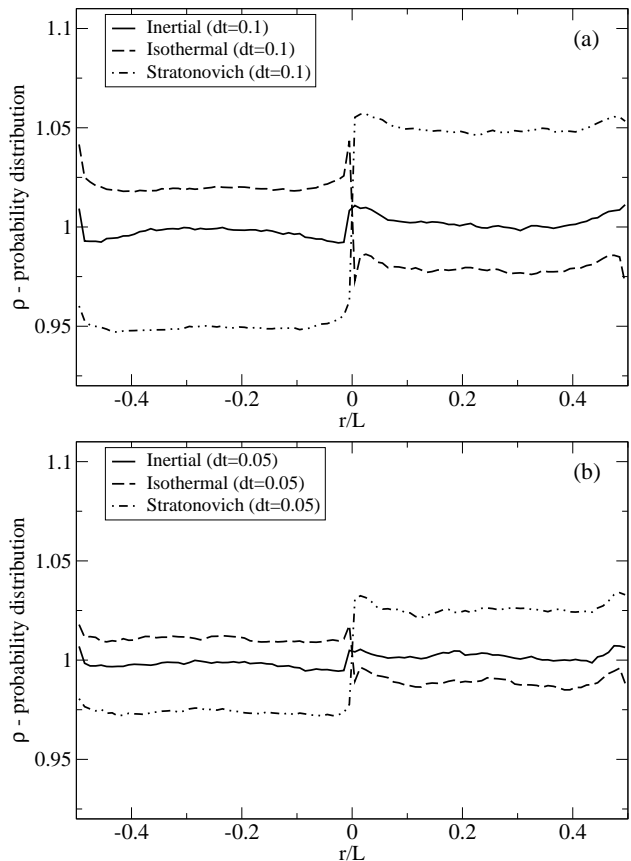}}
\end{center}
\vspace{-0.5cm}
\caption{Probability distribution computed using Stratonovich
(dashed-dotted), isothermal (dashed), and inertial (solid line)
interpretations.  The results correspond to a particle diffusing in a
flat potential with a step friction function. The results for the
different conventions were computed with (a) $dt=0.1$ and (b)
$dt=0.05$.}
\label{fig:step}
\end{figure}

The discussion in the previous paragraph is valid only for smooth
friction functions for which the change in the friction coefficient
during the time step is small. When $\alpha(r)$ exhibits rapid spatial
variations, the more physically-based inertial convention performs
much better than the isothermal one. This is nicely demonstrated in
Fig.~\ref{fig:step}, showing the results of simulations similar to
those depicted in Fig.~\ref{fig:smooth}, with the only difference
being that the sinusoidal friction function has been replaced with the
step-function $\alpha(r)=0.5+4.5\Theta(r)$, where $\Theta(r)$ is the
Heaviside step function. As in Fig.~\ref{fig:smooth}, the deviation
from a uniform probability distribution depicted in
Fig.~\ref{fig:step} follows the form of the simulated friction
function. In contrast to Fig.~\ref{fig:smooth}, the isothermal and
inertial conventions do not generate similar trajectories and
distribution functions. The latter, albeit not exact, is clearly
superior to the former. For comparison, the results of the
Stratonovich convention are also displayed. Notice (by comparing
Figs.~\ref{fig:step}(a) for $dt=0.1$ and (b) for $dt=0.05$) the
fundamental difference between the Stratonovich and isothermal results
which appear as step functions with an amplitude scaling linearly with
$dt$, and the inertial convention which, away from the discontinuity
in $\alpha$ (at $r=0$), recovers the correct value $\rho=1$.

\begin{figure}[t]
\begin{center}
\scalebox{0.365}{\centering \includegraphics{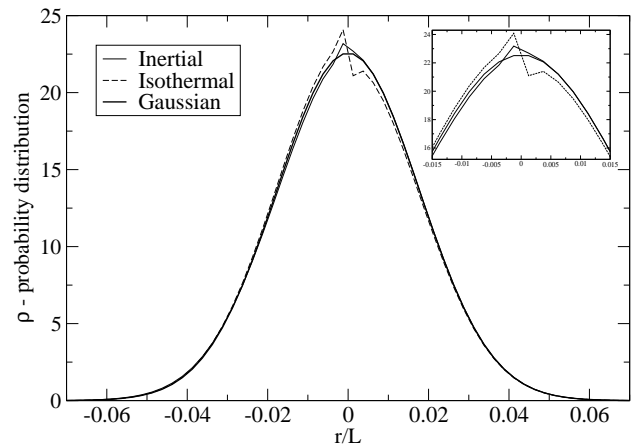}}
\end{center}
\vspace{-0.5cm}
\caption{Probability distribution computed using the isothermal
(dashed), and inertial (solid line) interpretations.  Results
correspond to a particle diffusing in a harmonic potential with
normalized natural frequency $\sqrt{k/m}=\sqrt{2}$ and with a step
friction function. Results for the different conventions were
computed with $dt=0.1$. Thick solid line depicts the exact
equilibrium Gaussian distribution. Inset shows a magnification of
the central region of the distribution.}
\label{fig:step_force}
\end{figure}

Fig.~\ref{fig:step_force} shows the distribution function computed
from simulations of a particle traveling in a medium with the same
step friction function as in Fig.~\ref{fig:step}, but in this case
within a harmonic potential well $U=kr^2/2$ with a normalized spring
constant $k=2$. When the deterministic force does not vanish (as in
this case), Eq.~(\ref{eq:alphat1}) can be modified to
$\bar{\alpha}_t\simeq\bar{\alpha}_r(r^n\to r^n+v^n dt+f^n dt^2/2m)$,
although the impact of the new term involving $f^n$ is nearly
negligible for small $dt$. Our results demonstrate, once again, the
advantage of the inertial interpretation over the isothermal one in
producing accurate distribution with relatively large time steps
($dt=0.1$ in Fig.~\ref{fig:step_force}). Notice that both
interpretations converge to the correct Gaussian form, $\rho_{\rm
eq}(r)=(k/2\pi k_BT)^{1/2}\exp(-kr^2/2k_BT)$, away from the interface
between the two friction regimes. This observation can be traced to
the fact that for a constant $\alpha$ and a harmonic potential, the
GJF integrator generates the exact Gaussian distribution \cite{gjf}.

\begin{figure}[t]
\begin{center}
\scalebox{0.35}{\centering \includegraphics{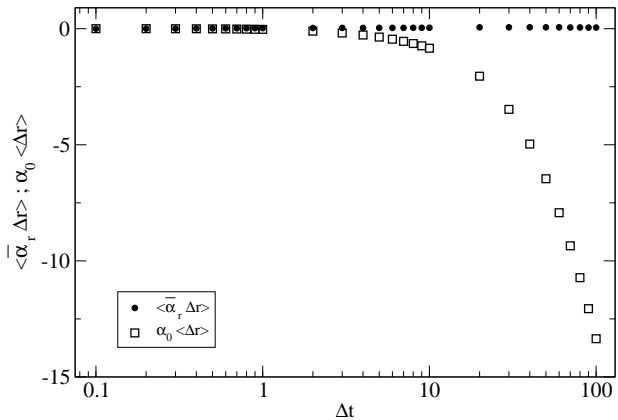}}
\end{center}
\vspace{-0.5cm}
\caption{Ensemble averages of $\bar{\alpha}_r\Delta r$ (solid
circles) and $\alpha_0\Delta r$ (open squares) as a function of time
$\Delta t$. Data computed from simulations of $10^7$ trajectories of a
particle traveling in a flat potential and a ramp friction function
Eq.~(\ref{eq:ramp}). The time step of the simulations: $dt=0.01$.}
\label{fig:drift}
\end{figure}

Having established the GJF integrator with the inertial convention as
the best available method for simulating dynamics in systems with
space-dependent friction, we now wish to use this method to examine
the validity of the theoretical predictions from section
\ref{sec:theory}. We start with the relation
$\langle\bar{\alpha}_r\Delta r\rangle=0$ (\ref{eq:drift}) governing
the drift of the particle in the absense of a deterministic force
($f=0$). We consider a particle moving in a medium with the following
``ramp'' friction function
\begin{eqnarray}
\alpha(r)=\left\{
\begin{array}{ll}
0.5 & {\rm for\ }r<-10 \\
0.5+0.225(r+10) & {\rm for\ }-10\leq r\leq 10 \\
5.0 & {\rm for\ }r>10
\end{array}
 \right..
\label{eq:ramp}
\end{eqnarray}
Starting at $r^0=0$ with a velocity randomly drawn from the
equilibrium Maxwell-Boltzmann distribution, we follow the particle and
record its position as a function of the time $\Delta t$. We set the
integration time step to $dt=0.01$, which is 10 times smaller than the
time step used in Figs.~\ref{fig:smooth}--\ref{fig:step_force}. The
ensemble average is calculated by repeating this procedure for $10^7$
different stochastic trajectories. The results, depicted by solid
circles in Fig.~\ref{fig:drift}, are in full agreement with
Eq.~(\ref{eq:drift}). For comparison, we also show (open squares)
the temporal dependence of $\alpha_0\langle\Delta r\rangle$ (where
$\alpha_0=\alpha(r^0)=2.75$). As expected, the data
reveals that there is an average drift toward negative values of $r$;
i.e., in the direction of the smaller friction coefficient.

\begin{figure}[t]
\begin{center}
\scalebox{0.35}{\centering \includegraphics{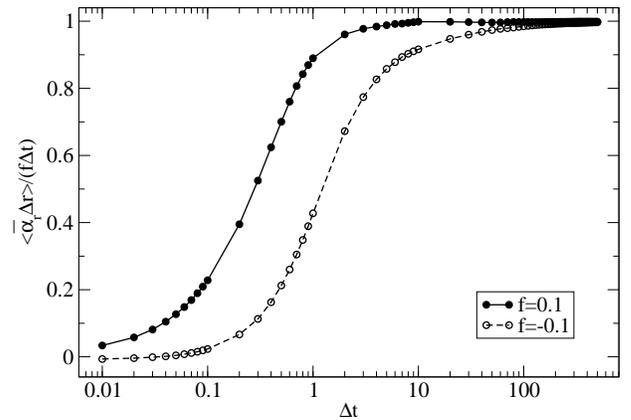}}
\end{center}
\vspace{-0.5cm}
\caption{Ratio between $\langle\bar{\alpha}_r\Delta r\rangle$ and
$f\Delta t$ as a function of time $\Delta t$ for $f=0.1$ (solid
circles, solid line) and $f=-0.1$ (open circles, dashed line). Data
computed from simulations of $10^7$ trajectories of a particle
traveling in a linear potential $-fr$ and a ramp friction function
Eq.~(\ref{eq:ramp}). The time step of the simulations: $dt=0.01$.}
\label{fig:force_drift}
\end{figure}

When $f\neq 0$, we expect the force-drift relationship $f\Delta
t=\langle\bar{\alpha}_r\Delta r\rangle$ Eq.~(\ref{eq:forcemes}) to hold
for large time scales $\Delta t\gg \tau$. To test the validity of this
prediction, we performed two sets of simulations similar to those
described in the previous paragraph, but now with non-vanishing forces
$f=0.1$ and $f=-0.1$. These values of $f$ have been chosen to make the
deterministic force comparable to the spurious force $f_s$
Eq.~(\ref{eq:spuriousf1}), and in order to examine both the situation
where $f$ and $f_s$ are parallel to each other as well as the case
where they point in opposite directions. The results of these
simulations are summarized in Fig.~\ref{fig:force_drift}, where we
plot the ratio $\langle \bar{\alpha}_r\Delta r\rangle/(f\Delta t)$ as
a function of $\Delta t$. The figure demonstrates that the ratio
indeed converges to unity at times much larger than the ballistic
relaxation time $\tau$, which can be evaluated by
$m/\max(\alpha)=0.2<\tau<2=m/\min(\alpha)$. The crossover into the
large $\Delta t$ regime occurs at somewhat smaller times when the
deterministic and spurious forces are opposite to each other.

\begin{figure}[t]
\begin{center}
\scalebox{0.35}{\centering \includegraphics{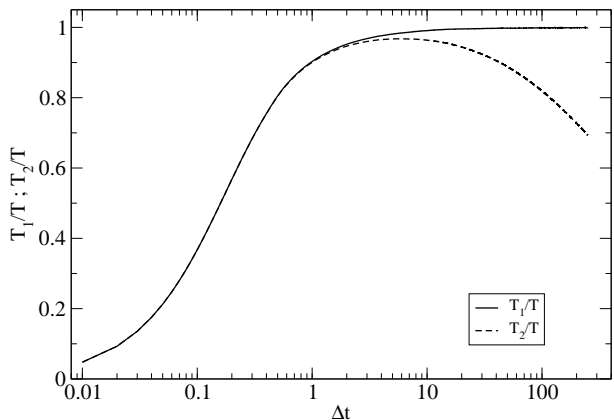}}
\end{center}
\vspace{-0.5cm}
\caption{Time dependence of the temperatures $T_1$
[Eq.~(\ref{eq:temp1}) - solid line] and $T_2$ [Eq.~(\ref{eq:temp2}) -
dashed line] defined, respectively, from the generalized (for
non-uniform $\alpha$) and standard (for constant $\alpha$) forms of
the fluctuation-dissipation relationship. Data computed from
simulations of $10^7$ trajectories of a particle traveling in a flat
potential and a parabolic friction function, where the initial
position is at the minimum of the parabola. The time step of the
simulations: $dt=0.01$.}
\label{fig:diffusion}
\end{figure}

Finally, we arrive at the generalized fluctuation-dissipation
relationship Eq.~(\ref{eq:fdrfinal}). To demonstrate the validity of
this equation, we consider the same particle with normalized mass
$m=1$ at constant temperature $T$, moving under the action of no force
($f=0$) in a medium with a parabolic friction function:
$\alpha(r)=10+0.1r^2$. At the initial time, an ensemble of $10^7$ such
(non-interacting) particles are placed at $r^0=0$, and with $dt=0.01$
we analyze their trajectories over time. The fact that the
trajectories start from the minimum of a parabolic friction function
ensures that, over time, the particles will arrive to further regions
of the system with an ever-increasing $\alpha(r)$, which would prevent
the friction coefficient from ``saturating''.  We define the
temperature $T_1$,
\begin{eqnarray}
k_BT_1=\frac{\left\langle \left(\bar{\alpha}_r\Delta
r\right)^2\right\rangle}{2\int_0^{\Delta t}
\langle\alpha(t)\rangle \, dt},
\label{eq:temp1}
\end{eqnarray}
which, according to Eq.~(\ref{eq:fdrfinal}), is expected to converge
to the thermodynamic temperature $T$ at large times $\Delta
t\gg\tau$. We also compare Eq.~(\ref{eq:fdrfinal}) with the standard
form of the fluctuation-dissipation relationship [see
Eqs.~(\ref{eq:fdr}) and (\ref{eq:diffusion})], featuring the
temperature $T_2$
\begin{eqnarray}
k_BT_2=\frac{\alpha_0\left\langle \left(\Delta
r\right)^2\right\rangle}{2\Delta t},
\label{eq:temp2}
\end{eqnarray}
that would have converged to unity had the friction coefficient been
constant $\alpha(r)=\alpha(r^0)=\alpha_0=10$. Our results, which are
summarized in Fig.~\ref{fig:diffusion}, demonstrate that $T_1$
indeed converges to $T$ - in full agreement with
Eq.~(\ref{eq:fdrfinal}). In contrast, the value of $T_2$ steadily
decreases at large times, which exemplifies that $\langle(\Delta
r)^2\rangle$ does not scale linearly with $\Delta t$ as suggested by
the conventional fluctuation-dissipation relationship. Notice that the
large time behavior of $T_2$ depends on the form of the function
$\alpha(r)$. In the case studied here, $\alpha(r)$ has a parabolic
form and the friction coefficient increases from the initial value of
$\alpha_0$.  This would naturally lead to a decrease in $T_2$, which
serves as a measure for the effective diffusion coefficient.

\section{Concluding remarks}
\label{sec:summary}
We conclude with the highlights of the study:

1.\ When a particle diffuses in a medium with spatially dependent
friction coefficient, it exhibits a drift toward the low-friction
end. The drift represents an inertial effect originating from the fact
that when the particle travels toward a less viscous side, it suffers
less dissipation and therefore travels longer distances. The drift
counters the tendency of the particle to get trapped, due to slower
diffusivity, in the more viscous parts of the system. The total amount
of time spent by the particle in each part of the system is,
obviously, independent of $\alpha(r)$ and depends only on the
potential energy (via the equilibrium distribution).

2.\ Since the drift results from an inertial effect, it needs to be
studied within the framework of the full Langevin equation
(\ref{eq:lang}) and not by using its overdamped (massless) limit
Eq.~(\ref{eq:browneq}). While the former equation of motion is simply
Newton's second law with friction and noise, the latter equation is
non-physical since it allows the velocity to diverge for an impulse of Gaussian
white noise. This leads to the ambiguity known as It\^{o}-Stratonovich
dilemma of the interpretation of the stochastic integral. The dilemma
is merely an artifact of the excessively simplified form
Eq.~(\ref{eq:browneq}). With the full Langevin equation, all the
conventions of assigning a value for the friction coefficient would
yield statistically similar trajectories in the limit when the time
step $dt\rightarrow0$. This leaves us with the ``lighter version'' of
the dilemma concerning the convention with the best rate of
convergence, which is an important computational issue.

3.\ We found the GJF integrator with the new inertial convention to be
the best method for Langevin dynamics simulations. The method produces
accurate thermodynamic behavior at large times for relatively large
integration time steps, even in systems with very rough friction
landscapes. The success of the method can be attributed in part to the
merits of the GJF integrator (which produces correct thermodynamic
response for constant $\alpha$ \cite{gjf}), and in part to the fact
that the inertial convention uses two different friction coefficients:
$\bar{\alpha}_r$ and $\bar{\alpha}_t$. The former of the two friction
coefficients governs the dissipative component of the integrated
Langevin equation (\ref{eq:intlang}), and the latter sets the
amplitude of the stochastic noise. While expression
(\ref{eq:rfriction}) for $\bar{\alpha}_r$ is exact, expression
(\ref{eq:tfriction}) for $\bar{\alpha}_t$ is not, but it ensures that
the requirement of Eq.~(\ref{eq:zeronoise}) for the noise is
satisfied. This requirement is rooted in the way that the random
collision forces are represented in the Langevin equation, where the
friction describes the mean force impulse, and the noise accounts for
the fluctuations around the mean force.

4.\ We derived three new equations to characterize the dynamics in
media with non-uniform friction.  Equations~(\ref{eq:drift}) and
(\ref{eq:fdrfinal}) describe the average and mean squared displacement
in the absence of a deterministic force ($f=0$), while
Eq.~(\ref{eq:forcemes}) gives the force-displacement relationship for
$f\neq 0$. Notice that only the first equation (\ref{eq:drift}) holds
for any time $\Delta t$, while the other two describe the asymptotic
behavior for $\Delta t\gg\tau$. The equations involve the variable
$\bar{\alpha}_r\Delta r=\Delta A$ [where $A(r)$ is the primitive
function of $\alpha_r(r)$], which emerges at the quantity that
characterizes the statistical properties of the dynamics. The validity
of the newly derived equations has been verified by computer
simulations.

5.\ We demonstrated that our generalized form of the
fluctuation-dissipation relationship (\ref{eq:fdrfinal}) is consistent
with Fick's second law (\ref{eq:ficklaw}), where the local diffusion
coefficient $D(r)=k_BT/\alpha(r)$. We reemphasize that diffusion is a
non-local process and, thus, $D(r)$ bears physical meaning only within
the context of a Focker-Planck differential equation, and only in
cases where $\alpha(r)$ is a smooth function. Notice that the
Focker-Planck differential equation can be (and, in fact, is usually)
derived from the overdamped limit of the Langevin equation of
motion. The agreement between our form of the fluctuation-dissipation
relationship and the Focker-Planck equation implies that ignoring
inertial effects does {\em not}\/ necessarily produce incorrect
equilibrium distributions. It simply means that the Focker-Planck
equation must be derived with care, i.e., with the appropriate
spurious drift term. The problem is the overdamped dynamics itself,
namely the attempt to calculate the physical trajectory of a particle
from a non-physical equation where its mass is set to zero.

This work was supported by the US Department of Energy Project
DE-NE0000536 000.

\end{document}